# Number of Near-Earth Objects and Formation of Lunar Craters over the Last Billion Years

**S. I. Ipatov**[a],*, **E. A. Feoktistova**[b], and **V. V. Svettsov**[c]

[a]*Vernadsky Institute of Geochemistry and Analytical Chemistry, Russian Academy of Sciences, Moscow, 119991 Russia*
[b]*Sternberg Astronomical Institute, Moscow State University, Moscow, 119992 Russia*
[c]*Institute of Geosphere Dynamics, Russian Academy of Sciences, Moscow, 119334 Russia*
*\*e-mail: siipatov@hotmail.com*

**Abstract**—We compare the number of lunar craters larger than 15 km across and younger than 1.1 Ga to the estimates of the number of craters that could have been formed for 1.1 Ga if the number of near-Earth objects and their orbital elements during that time were close to the corresponding current values. The comparison was performed for craters over the entire lunar surface and in the region of the Oceanus Procellarum and maria on the near side of the Moon. In these estimates, we used the values of collision probabilities of near- Earth objects with the Moon and the dependences of the crater diameters on the impactor sizes. According to the estimates made by different authors, the number density of known Copernican craters with diameters $D \geq 15$ km in mare regions is at least double the corresponding number for the remaining lunar surface. Our estimates do not contradict the growth in the number of near-Earth objects after probable catastrophic fragmentations of large main-belt asteroids, which may have occurred over the recent 300 Ma; however, they do not prove this increase. Particularly, they do not conflict with the inference made by Mazrouei et al. (2019) that 290 Ma ago the frequency of collisions of near-Earth asteroids with the Moon increased by 2.6 times. The number of Copernican lunar craters with diameters not smaller than 15 km is probably higher than that reported by Mazrouei et al. (2019). For a probability of a collision of an Earth-crossing object (ECO) with the Earth in a year equaled to $10^{-8}$, our estimates of the number of craters agree with the model, according to which the number densities of the 15-km Copernican craters for the whole lunar surface would have been the same as that for mare regions if the data by Losiak et al. (2015) for $D < 30$ km were as complete as those for $D > 30$ km. With this collision probability of ECOs with the Earth and for this model, the cratering rate may have been constant over the recent 1.1 Ga.

## INTRODUCTION

Small bodies are abundant in the Solar System. Some of them travel in the Solar System and may collide with planets and satellites, which results in craters. On the Moon and other atmosphereless bodies, craters persist for a long time (billions of years) since they are not exposed to an atmosphere and water. The preservation of craters on the Moon is also ensured by the fact that, over the past billion years, there have been no substantial geologic endogenic processes there (Bazilevskii et al., 1983). The impact cratering process was a subject of many studies (Gault et al., 1968; Hartmann, 1972; Melosh, 1989). Hartmann (1972) showed that the transition from simple to complex craters is determined by the gravity magnitude. Large craters with wall terraces, central peaks, and flat floors are classified as complex (Melosh, 1989). Some researchers (see, Kruger et al., 2015) proposed various values (from 15 to 20 km) for the crater diameter, under which the transition from simple to complex craters takes place on the Moon; moreover, this diameter may be different for mare and highland lunar regions (Kruger et al., 2018). For the terrestrial craters, this transition occurs at diameters of 3.2 km (Melosh, 1989). The impact cratering process is described in detail by Gault et al. (1968) and Melosh (1989); and this process is conventionally divided into three stages: contact, excavation, and modification. At the first stage, an impactor (a meteoroid, a comet, or an asteroid) impacts the surface of a target (a planet or a satellite). After contact, the excavation stage starts, which results in formation of the so-called "transient crater," the sizes of which substantially exceed the impactor size. The final stage of crater formation is modification. The process of modifying a transient crater and forming a complex one was considered in detail by Melosh and Ivanov (1999). Croft (1981, 1985) proposed the ratios between the diameters of a transient crater and a final complex crater for the Moon and the Earth.

To estimate the parameters of impactors from the characteristics of the formed craters, a series of scaling laws is used, i.e., the experimental data are extrapolated to the larger-scale events. Roddy et al. (1977) proposed the shock-explosion analogy method, which is based on substituting the energy of a charge by the kinetic energy of an impactor. Dienes and Walsh (1970) developed the late-stage equivalence principle, which introduces a scaling law for the diameter of a transient crater and the impactor size. This method was developed further by Holsapple and Schmidt (1980, 1982, 1987) and Schmidt and Housen (1987). The π-group scaling law, which is based on the analysis of dimensions of the physical quantities determining the cratering process and described in detail in the book by Melosh (1989), was proposed by Buckingham in 1914. This method was used in some studies to determine the ratio between the diameter of a transient crater and the characteristics of a target and an impactor (e.g., Pierazzo and Melosh, 2000; Ivanov et al., 2001); moreover, scaling relationships are widely used in web-based computer programs to estimate the crater sizes from the specified parameters of an impactor (Collins et al., 2005).

As was shown by Gilbert (1893) and Shoemaker (1962), the impact angles are randomly distributed with a maximum at 45°. According to the conclusion by Melosh (1989), the magnitude of the impact angle mostly influences the shape of the crater ejecta rather the crater itself. The ratio between the diameters of a crater and an impactor or the cratering effectiveness (the ratio of the mass ejected from a crater to the impactor mass) depending on the impact angle was analyzed by numerical modeling in several studies (e.g., Ivanov and Artemieva, 2001, 2002; Elbeshausen et al., 2009; Davison et al., 2011; Elbeshausen et al., 2013). Evidently, the cratering effectiveness decreases with decreasing the impact angle. From model calculations for the impact angles ranging from 30° to 90° (a vertical impact) and for an impact velocity of 6.5 km/s, Elbeshausen et al. (2009) found that the effectiveness of formation of a transient crater depends on the sine of the angle raised to some friction-dependent power. Under small impact angles, the transient crater may be elongated in shape; moreover, the threshold angle for transiting to elliptic-shaped craters goes down with decreasing target strength (Davison et al., 2011).

If the relation between the diameters of a crater and an impactor for intermediate values of the velocities and impact angles is known, the flux of bodies that form these craters may be determined from the observations, analysis, and counting of impact craters on the Moon. In the book *Asteroidno-Kometnaya Opastnost'* (2010) and the paper by Werner and Ivanov (2005), it is supposed that, during the last 3 Ga, the flux of crater-forming bodies has been almost stable, while about 4 Ga ago this flux was 100–500 times higher than the present one. The data on the lunar craters smaller than 100 m across suggested that the flux of crater-forming bodies has not changed for the last 100 Ma (*Asteroidno-Kometnaya Opastnost'*, 2010). Bottke et al. (2007) supposed that about 160 Ma ago the asteroid, fragments of which form the Baptistina family (the semimajor axis of asteroid 298 Baptistina is 2.26 AU), could have been destroyed in the asteroid belt, which induced the increase in the flux of bombarding bodies.

From the analysis of the ages of Copernican craters (i.e., the craters younger than 1.1 Ga), Mazrouei et al. (2019a) inferred that 290 Ma ago the frequency of col- lisions of near-Earth asteroids with the Moon increased by ~2.6 times. The estimates of the crater ages made by Mazrouei et al. (2019a) were based on the data of the Diviner radiometer onboard the *Lunar Reconnaissance Orbiter* (LRO) (i.e., on the analysis of the thermal characteristics of the material ejected in impacts). These authors also payed attention to the fact that the terrestrial craters with ages between 300 and 650 Ma are lacking and the older ones are almost completely absent. The twofold growth of the crater formation rate for the recent 300 Ma was also hypothesized earlier by McEwen et al. (1997) from the examination of bright ejecta rays originating in craters. These authors supposed that rayed craters on the far side of the Moon are younger than 1 Ga.

In the present paper we consider a probable change in the flux of crater-forming bodies for the recent billion years and discuss the probability of the increase in this flux for the recent 300 Ma. Our assessment is based on the comparison of the numbers of craters younger than 1.1 Ga estimated by Mazrouei et al. (2019a) and Losiak et al. (2015) and our estimates made under the assumption that the number and orbits of near-Earth objects have barely changed over the last 1.1 Ga. For our analysis, we use the data on the number of near-Earth objects, the estimates of probabilities of collisions of near-Earth objects with the Moon, and the dependences of the crater diameters on the impactor sizes.

Earlier, Ipatov et al. (2018, 2020) estimated the number of Copernican craters on the basis of the data by Losiak et al. (2011) and additionally considered the craters of unverified age and the first (i.e., best) integrity degree contained in the Morphology Catalog of Lunar Craters issued by the Sternberg Astronomical Institute (SAI) (Rodionova et al., 1987). Only one of these craters is in the table considered by Mazrouei et al. (2019a). Consequently, craters of this kind are not considered here.

When estimating the crater number, we used the catalog of lunar craters (Losiak et al., 2015) that includes the information on the age of craters from the papers by Wilhelms (1987) and Wilhelms and Bane (2009) and the data from the paper Mazrouei et al. (2019a). Wilhelms (1987) and Wilhelms and Bane (2009) estimated the crater ages from the statistics of craters and their degree of preservation. According to Losiak et al. (2015), only 66 craters with diameters $D \geq$ 10 km are dated as belonging to the Copernican period, and 38 of these are in mare regions of the near side of the Moon. On the far side of the Moon, there are fewer maria than on the near side; and they are smaller in size. The comparison of the data from the catalog by Losiak et al. (2015) to the distribution presented by the Neukum curve with the coefficients from Werner and Ivanov (2015, Eq. (21), Table 1) shows that these data for craters with $D$ < 30 km are incomplete (Fig. 1). To estimate the age of the craters formed in the last billion years, Mazrouei et al. (2015) used the data of the LRO Diviner radiometer. Mazrouei et al. (2015) proceeded from the assumption that the thermal properties of the material composing the crater ejecta depend on the crater age. In the younger impact structures, the ejecta material mostly contains large rock fragments, thermal inertia and, consequently, its temperature differs from the corresponding parameters of the surrounding material. The Diviner radiometer gathered the information on variations in the lunar surface temperature for a lunar day. Based on these data, Mazrouei et al. (2015) distinguished the craters younger than 1 Ga on the surface.

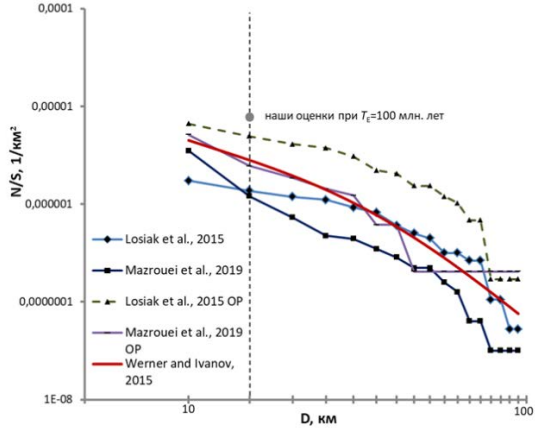

**Fig. 1.** The distribution of the number density $N/S$ of lunar craters younger than 1.1 Ga with diameters exceeding $D$ according to the data of Losiak et al. (2015), Mazrouei et al. (2019a), and Werner and Ivanov (2015). The data for the region in the Oceanus Procellarum and the other maria on the near side of the Moon are marked by OP. The rest of the data represent the entire lunar surface. A circle symbol shows our estimates of $N/S$ made with the characteristic time before a collision of an ECO with the Earth $T_E$ = 100 Ma (i.e., under the probability equal to $10^{-8}$ that an ECO will impact the Earth in a year). The vertical line corresponds to $D$ = 15 km; and the spaces between dashes on the abscissa correspond to 5 km.

Their results agree well with the distribution for Copernican craters considered by Werner and Ivanov (2015). However, in the paper by Mazrouei et al. (2019a), the number of Copernican craters was revised: according to the new results, the number of craters larger than 10 km across is only 111, and only 25 from them are in mare regions on the near side of the Moon. At the same time, Mazrouei et al. (2019a) note that the flux of large impactors probably increased ~260 Ma ago. They proposed that this increase may be caused by infalls of large fragments, which were formed due to destruction of one or several asteroids in the main belt, onto the lunar surface.

To estimate the age of craters, Ghent et al. (2014) and Mazrouei et al. (2019a) compared these craters to nine craters whose age was independently determined. Hergarten et al. (2019) questioned the estimates of the crater age obtained by Mazrouei et al. (2019a). Hergarten et al. (2019) noted that the observed facts may be interpreted in a different way and the data on nine craters used by Ghent et al. (2014) and Mazrouei et al. (2019a) may be insufficient for calibrating their model with the power law and disproving the exponential model. Mazrouei et al. (2019b) adduced arguments for their model. When analyzing the ages of craters, Kirchoff et al. (2013) concluded that, between 3 and 1 Ga, there might be long periods (longer than 600 Ma) with relatively small numbers of collisions, which gave place to shorter periods (about 200 Ma) of more intense bombardment.

For our estimates, we used the data on the number of Copernican craters for both the entire lunar surface and separately for the region of the Oceanus Procellarum and maria on the near side of the Moon. In the last section, we compare these data to our estimates of the number of craters, which could have been formed on the Moon for 1.1 Ga under the condition that, in this period, the distributions of ECOs over numbers, masses, densities, and elements of heliocentric orbits were on average the same as those currently observed. In the other sections, the data used in this comparison are discussed, specifically, the dependences of the crater diameters on the impactor sizes. The number of known Copernican craters and the data of the models were compared for several estimates of the collision probabilities of ECOs with the Earth (Werner and Ivanov, 2015; Emel'yanenko and Naroenkov, 2015; Granvik et al., 2018; Bottke et al., 1994; Dvorak and Pilat-Lohinger, 1999; Ipatov, 2000; Ipatov and Mather, 2004a). The ratio of the collision probabilities of near-Earth objects (NEOs) with the Earth to those with the Moon was assumed at 22 (Le Feuvre and Wieczorek, 2011).

## IMPACTOR DIAMETERS CORRESPONDING TO DIAMETERS OF LUNAR CRATERS

To determine the relation between the diameters of an impact crater and an impacting body, we used the well-known experiment-based relationship (Schmidt and Housen, 1987). For vertical impacts, it is written in the SI system as follows (Collins et al., 2005; Werner and Ivanov, 2015):

$$D_{tc} = 1.16\,(\rho_i/\rho_t)^{1/3}\,D_p^{0.78}\,U^{0.44}\,g^{-0.22}, \quad (1)$$

where $D_{tc}$ is the transient crater diameter, $\rho_i$ and $\rho_t$ are the densities of an impactor and a target, respectively, $D_p$ is the impactor (projectile) diameter, $U$ is the impact velocity, and $g$ the acceleration due to gravity. To determine the diameter of a final complex crater $D_v$ (by a rim), the following relationship may be used (Croft, 1981, 1985; Werner and Ivanov, 2015)

$$D_v = D_{sc}^{-0.18} D_{tc}^{1.18}, \quad (2)$$

where $D_{sc}$ is the crater diameter corresponding to the transition from simple to complex craters. This diameter is approximately 15 and 3.2 km for the Moon and the Earth, respectively (Melosh, 1989). We note that the inaccuracy in determining $D_{sc}$ caused by a small power exponent, equaled to 0.18, in formula (2) yields an error of only several per cent. Taking into consideration Eq. (1), formula (2) may be transformed to

$$D_v = 1.18\,D_{sc}^{-0.18}\,(\rho_i/\rho_t)^{0.393}\,D_p^{0.92}\,U^{0.52}\,g^{-0.26} \quad (3)$$

Table 1. Sizes and coordinates of craters younger than 1.1 Ga according to Losiak et al. (2015). The craters located in the region of the Oceanus Procellarum and maria on the near side of the Moon are asterisked

| No. | Name | Latitude | Longitude | Diameter, km | Depth, km | Depth–diameter ratio | Age, Ga |
|---|---|---|---|---|---|---|---|
| 1 | Vavilov | −0.9 | −138.8 | 98.2 | 6 | 0.061 | ≤1.1 |
| 2 | Copernicus* | 9.6 | −20.1 | 96.1 | 3 | 0.031 | 0.8 |
| 3 | Hayn | 64.6 | 83.9 | 86.2 | 3.6 | 0.042 | ≤1.1 |
| 4 | Tycho | −43.3 | −11.2 | 85.3 | 4.5 | 0.052 | ≤1.1 |
| 5 | King | 5 | 120.5 | 76.2 | 3.8 | 0.05 | ≤1.1 |
| 6 | Stevinus* | −32.5 | 54.1 | 71.5 | 3.8 | 0.053 | ≤1.1 |
| 7 | Jackson | 22.1 | −163.3 | 71.4 | 4.7 | 0.066 | ≤1.1 |
| 8 | Philolaus* | 72.2 | −32.9 | 71.4 | 3.8 | 0.053 | 1.02 |
| 9 | O'Day | −30.4 | 157.3 | 70.4 | 3.8 | 0.054 | ≤1.1 |
| 10 | Eudoxus* | 44.3 | 16.2 | 70.2 | 4.5 | 0.064 | ≤1.1 |
| 11 | Zucchius* | −61.4 | −50.7 | 63.2 | 3.6 | 0.057 | ≤1.1 |
| 12 | Ohm | 18.3 | −113.8 | 61.8 | 4.5 | 0.072 | ≤1.1 |
| 13 | Carpenter* | 69.5 | −51.2 | 59.1 | 3.8 | 0.064 | ≤1.1 |
| 14 | Taruntius* | 5.5 | 46.5 | 57.3 | 1.5 | 0.026 | ≤1.1 |
| 15 | Aristillus* | 33.9 | 1.2 | 54.4 | 3.85 | 0.071 | 1.1 |
| 16 | Anaxagoras* | 73.5 | −10.2 | 51.9 | 4.2 | 0.081 | ≤1.1 |
| 17 | Rutherford | −61.2 | −12.3 | 50 | 3 | 0.06 | ≤1.1 |
| 18 | Crookes | −10.4 | −165.1 | 48.3 | 4.5 | 0.093 | ≤1.1 |
| 19 | Bel'kovich | 63.6 | 93.6 | 47 | 2 | 0.042 | ≤1.1 |
| 20 | Wiener F | 41.2 | 150 | 44.9 | 3.3 | 0.073 | ≤1.1 |
| 21 | Bürg | 45.1 | 28.2 | 41 | 3.25 | 0.079 | ≤1.1 |
| 22 | Glushko* | 8.1 | −77.7 | 40.1 | 2.2 | 0.051 | ≤1.1 |
| 23 | Aristarchus* | 23.7 | −47.5 | 40 | 3.2 | 0.08 | 0.13–0.18 |
| 24 | Harpalus* | 52.7 | −43.5 | 39.8 | 3.4 | 0.085 | 0.26 |
| 25 | Virtanen | 15.6 | 176.7 | 39.7 | 3.3 | 0.083 | ≤1.1 |
| 26 | Autolycus* | 30.7 | 1.5 | 38.9 | 3.7 | 0.095 | ≤1.1 |
| 27 | Palitzsch B | −26.4 | 68.4 | 37.9 | 3.1 | 0.082 | ≤1.1 |
| 28 | Guthnick | −47.8 | −94 | 37 | 4 | 0.11 | ≤1.1 |
| 29 | Necho | −5.3 | 123.2 | 36.9 | 3.8 | 0.10 | ≤1.1 |
| 30 | Milne H | −35.8 | 111.3 | 36.8 | 3.1 | 0.084 | ≤1.1 |
| 31 | Das | −26.5 | −137.1 | 36 | 3.5 | 0.097 | ≤1.1 |
| 32 | Godin* | 1.8 | 10.2 | 34.3 | 3.2 | 0.093 | ≤1.1 |
| 33 | Gassendi A* | −15.6 | −39.8 | 32.2 | 3 | 0.093 | ≤1.1 |
| 34 | Petavius B* | −19.9 | 57 | 32 | 3 | 0.094 | ≤1.1 |
| 35 | Thales* | 61.7 | 50.3 | 30.8 | 3.5 | 0.114 | ≤1.1 |
| 36 | Kepler* | 8.1 | −38 | 29.5 | 2.9 | 0.098 | 0.625–1.25 |
| 37 | Schomberger A | −78.6 | 23.5 | 29.4 | 2.9 | 0.099 | ≤1.1 |
| 38 | Birkhoff Z | 61 | −145.9 | 28.6 | 2.9 | 0.101 | ≤1.1 |
| 39 | Faraday C | −43.3 | 8 | 28.5 | 2.9 | 0.102 | ≤1.1 |
| 40 | Mädler* | −11 | 29.8 | 27.6 | 3 | 0.11 | ≤1.1 |
| 41 | Proclus* | 16.1 | 46.9 | 26.9 | 4 | 0.143 | ≤1.1 |
| 42 | Triesnecker* | 4.2 | 3.6 | 25 | 2.7 | 0.108 | ≤1.1 |
| 43 | Mösting* | −0.7 | −5.9 | 24.4 | 2.9 | 0.12 | ≤1.1 |
| 44 | Giordano Bruno | 36 | 102.9 | 22.1 | 2.7 | 0.122 | ≤1.1 |

Table 1. (Contd.)

| No. | Name | Latitude | Longitude | Diameter, km | Depth, km | Depth–diameter ratio | Age, Ga |
|---|---|---|---|---|---|---|---|
| 45 | Conon* | 21.7 | 2 | 21 | 2.7 | 0.128 | ≤1.1 |
| 46 | Thebit A* | −21.6 | −4.9 | 19.9 | 2.6 | 0.131 | ≤1.1 |
| 47 | Lichtenberg* | 31.9 | −67.7 | 19.5 | 2.6 | 0.133 | ≤1.1 |
| 48 | Pytheas* | 20.6 | −20.6 | 18.8 | 2.5 | 0.133 | ≤1.1 |
| 49 | Dawes* | 17.2 | 26.3 | 18 | 2.2 | 0.122 | ≤1.1 |
| 50 | Bellot* | −12.5 | 48.2 | 17.5 | 3 | 0.17 | ≤1.1 |
| 51 | Dionysius* | 2.8 | 17.3 | 17.3 | 3 | 0.158 | ≤1.1 |
| 52 | Rutherford | 10.6 | 137.1 | 16 | 2 | 0.125 | ≤1.1 |
| 53 | Janssen K | −46.2 | 42.3 | 15 | 2.4 | 0.16 | ≤1.1 |
| 54 | Messier* | −1.9 | 47.7 | 13.8 | 2.2 | 0.16 | ≤1.1 |
| 55 | Tharp | −30.6 | 145.6 | 13.5 | 2.2 | 0.162 | ≤1.1 |
| 56 | Milichius | 10 | −30.2 | 12.2 | 3 | 0.245 | ≤1.1 |
| 57 | Gambart A* | 1 | −18.8 | 11.6 | 2.3 | 0.198 | ≤1.1 |
| 58 | Sulpicius Gallus* | 19.6 | 11.7 | 11.6 | 2.3 | 0.198 | ≤1.1 |
| 59 | Furnerius A | −33.5 | 59 | 11.2 | 2.3 | 0.205 | ≤1.1 |
| 60 | Petavius C* | −27.7 | 60 | 11.2 | 2.2 | 0.198 | ≤1.1 |
| 61 | Goddard A* | 17.1 | 89.7 | 11.1 | 2.2 | 0.202 | ≤1.1 |
| 62 | Cameron* | 6.2 | 45.9 | 10.9 | 2.2 | 0.202 | ≤1.1 |
| 63 | Encke B* | 2.4 | −36.8 | 10.9 | 2.1 | 0.196 | ≤1.1 |
| 64 | Kepler A* | 7.1 | −36.1 | 10.7 | 2.1 | 0.198 | ≤1.1 |
| 65 | Fraunhofer C | −38.5 | 58.4 | 10.6 | 2.1 | 0.21 | ≤1.1 |
| 66 | Bessarion* | 14.9 | −37.3 | 10 | 2.2 | 0.198 | ≤1.1 |

For complex craters larger than 15 km across and by assuming $D_{sc}$ = 15 km, the gravity acceleration corresponding to that of the Moon, and equal values for the densities of an impactor and a target, it is easy to find from relationship (3) that the diameter of a final crater produced by a vertical impact Dv is (expressed in kilometers)

$$D_v = 3.94\,D_p^{0.92} \cdot U^{0.52} = 3.94\,(D_p \cdot U^{0.57})^{0.92}, \quad (4)$$

where $D_p$ is the impactor (projectile) diameter expressed in kilometers and $U$ is the impact velocity expressed in kilometers per second. The crater size also depends on the impact angle θ (between the velocity vector and the surface) as $(\sin\theta)^{1/3}$ (Melosh, 1989; Collins et al., 2005). Since the probability of the impact at an angle θ is sin2θ, to determine the crater diameter averaged over angles, the product of expression (4) and $(\sin\theta)^{1/3}\sin2\theta$ should be integrated with respect to the angles ranging from 0° to 90°. This will yield the expression analogous to Eq. (4), where a factor of 3.94 is substituted by 3.2:

$$D_v = 3.2\,D_p^{0.92} \cdot U^{0.52} = 3.2\,(D_p \cdot U^{0.57})^{0.92}. \quad (5)$$

The quantity $D_v$ is most sensitive to $U$. Since $U^{0.57 \times 0.92}$ is close to $U^{0.5}$, $D_v$ changes approximately twofold if

$U$ changes fourfold. Minton et al. (2015, Fig. 11) show that the ratio for the diameters of a crater and an impactor $D_c/D_p$ varies by more than 10 times under the same value of $D_p$ and different possible velocities of impacts.

Minton et al. (2015) assumed that the root-mean-square velocity of collisions between the Moon and the former main-belt asteroids migrated to the Earth's orbit is 18.3 km/s. However, the collision velocities shown by Minton et al. (2015, Fig. 10) are in a range of 7 to 50 km/s. Le Feuvre and Wieczorek (2011) and Stuart and Binzel (2004) determined the mean velocity of collisions with the Moon as 19.3 and 19.7 km/s, respectively. It is believed that a main source of near-Earth objects is the asteroid belt. However, some NEOs come from the regions that are distant from the Sun. For the bodies coming from beyond Jupiter's orbit and falling onto the Earth, the typical velocities of collisions with the Earth are somewhat larger than those for the NEOs originating in the asteroid belt. According to estimates by Marov and Ipatov (2018), the typical velocities of these bodies, which entered the Earth's activity sphere, are in a range of 21–24 km/s. Ipatov (2000) noted that, among the bodies that came from zones of the giant planets, the portion of bodies with Earth-crossing orbits is an order of magnitude higher than that of bodies with Mars-crossing orbits and their eccentricities usually exceed 0.6. Since it is not the case for known asteroids, these results suggest that most Amor asteroids should have come from the asteroid belt rather than beyond Jupiter's orbit. Asteroids are also considered as a main source of NEOs in many other papers of different authors. Granvik et al. (2018) believe that more than 90% of NEOs originated from asteroids.

From formula (5) we obtain that the mean diameter of a crater produced by impactors with a diameter of $D_p = 1$ km is $D_V = 14.7$, 15.3, and 16.9 km for impact velocities of 18.3, 19.7, and 24 km/s, respectively. In other words, the scattering in the $D_V$ values for this range of velocities is relatively small. Below, we will use $D_V = 15$ km.

For the Earth, Werner and Ivanov (2015) derived the formula close to Eq. (5), but containing a coefficient of 4 and somewhat differing exponents (0.58 instead of 0.57, and 0.91 instead of 0.92). This formula is based on numerical calculations, for which a number of assumptions were made. The formation of craters produced by vertical impacts on the Moon was numerically simulated by Silber et al. (2017) for several values of collision velocities. In these calculations, it was assumed that the impactor is composed of dunite with a density of 3.32 g/cm³, while the target material is granite with a density of 2.63 g/cm³. Formula (4) for vertical impacts was derived for equal values of the impactor and target densities. Consequently, when comparing to the results of calculations by Silber et al. (2017), Eq. (4) should be multiplied by the density ratio 3.32/2.63 exponentiated by 0.393. This enhances the estimate of the crater diameter by 10%. Finally, it turns out that, for complex craters, the result of Eq. (4) corrected for the density difference diverges from the result of Silber et al. (2017) by less than 5%.

Let us compare the crater diameters for the Moon and the Earth obtained with formula (3). The factor is $D_{sc}^{-0.18} \approx 0.757$, while $g^{0.26} \approx 1.60$. Their product is 1.2. In other words, if the densities of the Earth and the Moon are the same (see the citations below) and the impactor parameters are also the same, the crater on the Moon is 1.2 times larger in diameter than that on the Earth. In particular, if the impact on the Earth produces a crater 15 km across, the same impact on the Moon produces a crater 18 km across. The mean density of rocks in the Earth crust is 2.7 g/cm³ (Koronovskii, 2018). The density of the lunar crust varies in different regions, from 2.3 to 2.9 g/cm³ according to the Gravity Recovery and Interior Laboratory (GRAIL) data, while its average for the lunar highlands is 2.55 g/cm³ (Wieczorek et al., 2013).

## THE NUMBER OF LUNAR CRATERS WITH DIAMETERS LARGER THAN 15 KM

We analyzed the number of lunar craters larger than 15 km in diameter and younger than 1.1 Ga (i.e., dated to the Copernican period of the geologic history of the Moon) on the whole lunar surface and in the region of the Oceanus Procellarum, Mare Imbrium, Mare Tranquillitatis, and other maria on the near side of the Moon. The Oceanus Procellarum and the other maria on the near side of the Moon are plains and impact basins filled with molten basalt lava. The ratio of the considered area to that of the entire lunar surface is $r_{reg} = 0.155$. This is a corrected value, while Ipatov et al. (2018, 2020) used a ratio of 0.176 earlier.

Below, we use, primarily, the data by Mazrouei et al. (2019a). In that paper, the estimates of the crater ages were based on the LRO Diviner radiometer data, which allowed the thermal properties of the lunar surface in the vicinity of the investigated craters to be analyzed. As distinct from the earlier studies, Mazrouei et al. (2019a) considered the data obtained by the radiometer right up to 2012.

The catalog of lunar craters (Losiak et al., 2015) contains the information on the crater ages based on the papers by Wilhelms (1987) and Wilhelms and Bane (2009), who used traditional methods to date craters. As has been shown above in the analysis of Fig. 1, for the craters with diameters $D < 30$ km, the information in the catalog of lunar craters (Losiak et al., 2015) is not complete. However, when discussing some issues, we considered the data from that paper as well; which was due to the following reasons. The data of Losiak et al. (2015) exhibit the smallest estimate for the number of craters with diameters $D < 30$ km, which, even for $15 \le D \le 30$ km, turns out to be higher than the data by Mazrouei et al. (2019a) with a correct slope of the Neukum curve. Together with the data of Werner and Ivanov (2015) and our estimates of the crater number given below, this allows us to suppose that, in Fig. 1, the curve for the data of Losiak et al. (2015) should lie higher at $D < 30$ km. We compared the crater number density for the entire lunar surface and the mare region from the data by both groups, Mazrouei et al. (2019a) and Losiak et al. (2015), and found the differences in the crater number densities between maria and highlands in both papers. The data of Losiak et al. (2015) are also used below to compare the depth-to-diameter ratios for craters in maria and highlands and to compare the number of craters, particularly with diameters larger than 30 km, to the data by Mazrouei et al. (2019a).

Table 2. Sizes, coordinates, and ages of Copernican craters located in the region of the Oceanus Procellarum and maria on the near side of the Moon according to Mazrouei et al. (2019a)

| No. | Name | Latitude, deg | Longitude, deg | Diameter, km | Depth, km | Depth–diameter ratio | Age, Ga |
|---|---|---|---|---|---|---|---|
| 1 | Copernicus | 10 | –20 | 97 | 3 | 0.031 | 0.797 |
| 2 | Glushko | 8 | –78 | 43 | 2.2 | 0.051 | 0.196 |
| 3 | Aristarchus | 24 | –47 | 40 | 3.2 | 0.080 | 0.164 |
| 4 | Petavius B | –27 | 57 | 34 | 3 | 0.088 | 0.224 |
| 5 | Thales | –20 | 50 | 32 | 2.5 | 0.078 | 0.061 |
| 6 | Kepler | 8 | –38 | 32 | 2.7 | 0.084 | 0.93 |
| 7 | Proclus | 16 | 47 | 28 | 4 | 0.143 | 0.253 |
| 8 | Lalande | –4 | –9 | 24 | 2.9 | 0.121 | 0.495 |
| 9 | Schwabe F | 66 | 50 | 21 | 2.8 | 0.133 | 0.814 |
| 10 | Dionysius | 3 | 17 | 19 | 3 | 0.158 | 0.492 |
| 11 | Dawes | 17 | 26 | 18 | 2.2 | 0.122 | 0.454 |
| 12 | Carrel | 11 | 27 | 17 | 2.2 | 0.129 | 0.295 |
| 13 | Sirsalis F | –14 | 300 | 13 | 2.5 | 0.192 | 0.629 |
| 14 | Darney C | –14 | 334 | 13 | 2.5 | 0.192 | 0.582 |
| 15 | Egede A | 52 | 11 | 13 | 2.6 | 0.20 | 0.084 |
| 16 | Mösting A | –3 | 335 | 13 | 2.9 | 0.223 | 0.113 |
| 17 | Gambart A | 1 | 341 | 12 | 2.6 | 0.217 | 0.344 |
| 18 | Cameron | 6 | 46 | 11 | 2.2 | 0.20 | 0.48 |
| 19 | Alfraganus C | –6 | 18 | 11 | 2.5 | 0.227 | 0.433 |
| 20 | Messier A | –2 | 47 | 11 | 2.4 | 0.218 | 0.128 |
| 21 | Aratus | 24 | 5 | 11 | 1.9 | 0.173 | 0.421 |
| 22 | Euclides C | –13 | 330 | 11 | 2 | 0.182 | 0.137 |
| 23 | Arnold G | 67 | 31 | 10 | 2.2 | 0.22 | 0.223 |
| 24 | Bessarion | 15 | 323 | 10 | 2.1 | 0.21 | 0.164 |
| 25 | Democritus A | 62 | 32 | 10 | 2 | 0.20 | 0.218 |

Tables 1 and 2 contain information from the papers by Mazrouei et al. (2019a) and Losiak et al. (2015) about the names, coordinates (latitudes and longitudes), diameters, and depths of the craters, the age of which does not exceed 1.1 Ga. For some craters from Table 1 and all craters from Table 2, the estimates of their ages (expressed in Ga) are also shown. We use the data presented in Tables 1 and 2 to analyze the temporal changes in the number of near-Earth objects for estimating the number of craters larger than 15 or 18 km across (in these estimates, the diameter values were rounded to the nearest whole number) and to study the dependence of the diameter-to-depth ratio of craters on their diameter and the dependence of the number of craters on their diameter. The estimates of the number of craters with diameters $D > 18$ km are used below to analyze the data of Table 3 and consider the differences between the numbers of craters with $15 < D < 18$ km reported by Mazrouei et al. (2019a) and Losiak et al. (2015).

According to Losiak et al. (2015), 53 craters smaller than 15 km across are dated to the Copernican period; and 29 from them are located in the mare region on the near side of the Moon (Table 1). According to Mazrouei et al. (2019a), the number of craters of this kind is 44 and 22, respectively (Table 2).

According to the data from Table 1, which are based on the papers by Wilhelms (1987) and Wilhelms and Bane (2009), in the regions of the Oceanus Procellarum and the other maria on the near side of the Moon (which comprises 0.155 of the lunar surface area), the number of craters not smaller than 18 and 15 km in diameter is 27 and 29, respectively. According to the data of Table 2 obtained from the paper by Mazrouei et al. (2019a, Supplementary material, Table S1), the regions of the Oceanus Procellarum and the other maria on the near side of the Moon contain 11, 12, and 16 craters with diameters not smaller than 13, 17, and 18 km, respectively. That is, Table 2 contains only one crater for an interval of $13 < D < 17$ km and four craters for a much narrower interval of $17 < D < 18$ km.

Table 3. The number of craters younger than 1.1 Ga with diameters not smaller than 15, 18, and 30 km—$N_{reg15}$, $N_{reg18}$, and $N_{reg30}$, respectively. These values were obtained for maria on the near side of the Moon, the entire lunar surface, and highlands (i.e., the difference between the areas of the entire lunar surface and the maria on the near side) from the data by Mazrouei et al. (2019a) and Losiak et al. (2015). The ratio $r_{reg}$ of the considered area to that of the Moon is 0.155, 1.0, and 0.845 for the maria on the near side, all craters, and the highlands, respectively

|  | Maria on the near side (Mazrouei et al., 2019a) | Highlands (Mazrouei et al., 2019a) | Moon as a whole (Mazrouei et al., 2019a) | Maria on the near side (Losiak et al., 2015) | Highlands (Losiak et al., 2015) | Moon as a whole (Losiak et al., 2015) |
|---|---|---|---|---|---|---|
| $r_{reg}$ | 0.155 | 0.845 | 1.0 | 0.155 | 0.845 | 1.0 |
| $N_{reg15}$ | 12 | 32 | 44 | 29 | 24 | 53 |
| $N_{reg18}$ | 11 | 20 | 31 | 27 | 22 | 49 |
| $N_{reg30}$ | 6 | 8 | 14 | 17 |  | 35 |
| $N_{reg15}/r_{reg}$ | 77 | 38 (77/38 ≈ 2.0) | 44 (77/44 ≈ 1.75) | 187 | 28 (187/28 ≈ 6.8) | 53 (187/53 ≈ 3.5) |
| $N_{reg18}/r_{reg}$ | 71 | 24 (71/24 ≈ 2.96) | 31 (71/31 ≈ 2.3) | 174 | 26 (174/26 ≈ 6.7) | 49 (174/49 ≈ 3.55) |
| $N_{reg30}/r_{reg}$ | 39 | 9.5 (39/9.5 ≈ 4.1) | 14 (39/14 ≈ 2.8) | 110 | 18 (110/18 ≈ 6.1) | 35 (110/35 ≈ 3.14) |

The ratios of the depths of craters to their diameters listed in Tables 1 and 2 take mostly larger values for smaller crater diameters. However, even for roughly the same diameters, the depths may differ twofold (e.g., compare lines 17–19 for highland craters and lines 13 and 14 for mare craters in Table 1).

Table 3 contains the numbers of craters younger than 1.1 Ga with diameters not smaller than 15, 18, and 30 km—$N_{reg15}$, $N_{reg18}$, and $N_{reg30}$, respectively. These numbers were obtained from the data by Mazrouei et al. (2019a) and Losiak et al. (2015) for the maria on the near side of the Moon and the entire lunar surface. The ratio $r_{reg}$ for the considered area to that for the Moon is 0.155 and 0.985 for the maria on the near side and all craters considered by Mazrouei et al. (2019a), respectively. Mazrouei et al. (2019a) examined the region between 80° N and 80° S. Since the difference between 0.985 and 1.0 produces almost no effect on the quantities contained in Table 3, we show there the data for $r_{reg} = 1$ instead of those for $r_{reg} = 0.985$. In Table 3, there are also the ratios $N_{reg15}/r_{reg}$, $N_{reg18}/r_{reg}$, and $N_{reg30}/r_{reg}$, which are the estimates of the number of craters on the whole lunar surface under the condition that the number density of craters is the same as that for the considered mare region.

For the entire lunar surface, the number of craters with diameters from 15 to 18 km obtained by Losiak et al. (2015) and Mazrouei et al. (2019a) is four and 13 (differing by more than three times), respectively; the corresponding estimates for the maria are four and one. This difference once again characterizes the incompleteness of the data for craters with $D < 30$ km reported by Losiak et al. (2015), which has been noted in the analysis of Table 1 (see Introduction).

The values of $N_{reg15}/r_{reg}$, $N_{reg18}/r_{reg}$, and $N_{reg30}/r_{reg}$ listed in Table 3 were obtained from the extrapolation based on the data for the region of the Oceanus Procellarum and other maria on the near side of the Moon; they are $k_{OP}$ times larger ($k_{OP} > 1.7$) than the values obtained for the craters on the whole, or almost whole, surface of the Moon. For $N_{reg15}/r_{reg}$, $k_{OP} = 1.75$ and 3.53 for the data by Mazrouei et al. (2019a) and Losiak et al. (2015), respectively. For $N_{reg18}/r_{reg}$, the analogous values of $k_{OP}$ are 2.29 and 3.55; while they are 2.79 and 3.14 for $N_{reg30}/r_{reg}$. Thus, the number density of the craters with diameters $D > 15$ km and $D > 30$ km, which can be dated as Copernican, in the mare regions on the near side of the Moon are larger than the corresponding quantities for the entire lunar surface. Moreover, the number density ratio $k_{OP}$ for craters with $D \geq 15$ km on the mare to that on the entire lunar surface according to Losiak et al. (2015) is larger than that from the data by Mazrouei et al. (2019a) by 2.35 times (i.e., 3.53/1.75); the corresponding difference in $k_{OP}$ for $D \geq 30$ km is ~1.13 times (i.e., 3.14/2.79). In other words, the $k_{OP}$ values derived from these papers for $D \geq 30$ km do not differ much, though the number of such craters examined by Mazrouei et al. (2019a) is ~2.5 times smaller (i.e., 35/14) than that from the paper by Losiak et al. (2015). Apparently, the data by Mazrouei et al. (2019a) underestimate the number of Copernican craters in the whole range of diameters. The ratio of the crater number densities for highlands and oceans is even higher than the analogous ratio $k_{OP}$ for the whole lunar surface and oceans. For craters larger than 15 and 30 km across, the ratio of the crater number densities for maria is higher than that for highlands by seven and six times, respectively, according to Losiak et al. (2015) and by two and four times, respectively, according to Mazrouei et al. (2019a).

## PROBABILITIES OF COLLISIONS OF NEAR-EARTH OBJECTS WITH THE MOON

According to the data of October 26, 2019 (https://cneos.jpl.nasa.gov/stats/size.html), the number of detected NEOs, i.e., the objects with the perihelion distances smaller than 1.3 AU, with the diameters $d > 1$ km, was 900, while the total number $N_1$ of NEOs is estimated as 920. This estimate of the number of NEOs available online is actually related to the number of NEOs with the absolute magnitude $H$ smaller than 17.75. The diameters of NEOs with $H = 17.75$ are 1 km if the albedo is 0.14. In reality the albedo of NEOs may differ, and the number of NEOs larger than 1 km across may differ from that with $H \leq 17.75$. Granvik et al. (2018) estimated the number of NEOs with $H \leq 17.75$ as $962^{+52}_{-56}$. According to their analysis, 58% of 962 NEOs cross the Earth's orbit. The albedo of S-type asteroids in the main belt is $0.208 \pm 0.079$, while this quantity is $0.071 \pm 0.040$ for C-type asteroids (Usui et al., 2013). We specify here probable ranges of the albedo variations. The densities of asteroids also vary. According to Carry (2012), the density of S- and C-type asteroids is $2.66 \pm 1.29$ and $1.57 \pm 1.38$ g/cm$^3$, respectively. The diameters of asteroids with the intermediate albedo values and the magnitude $H = 17.75$ are 0.82 and 1.4 km for the S and C type, respectively. If the density of the target is assumed as that for S-type asteroids (which is correct, since the density of S-type asteroids is close to that of the lunar crust), these asteroids with a magnitude of $H = 17.75$ produce, on average, the craters 12.5 km in diameter (if the impact velocity is 19 km/s), while the corresponding crater diameter is $D = 15.7$ km for C-type asteroids. Since C-type asteroids make up the majority (~75%), the number of asteroids with $H \leq 17.75$ should statistically correspond well to the number of craters larger than 15 km across. This number of craters larger than 15 km across can be somewhat smaller at the expense of S-type asteroids. Sporadic fragmentations of large asteroids should result in increasing the number of craters regardless the albedo of fragments. However, the temporal variations in the contribution of different sources of NEOs may slightly change the characteristic dependence of the crater diameter on the impactor diameter.

Ipatov and Mather (2004a, 200b) estimated the ratio of the number of NEOs to the number of Earth-crossing objects (ECOs) $k_{ECO} = 1300/756 \approx 1.72$. The analogous ratio was calculated for the objects detected before June 1, 2010 (6718/3906 ≈ 1.72) (*Asteroidno-Kometnaya Opasnost'*, 2010). As follows from the NASA data by August 2019 (https://cneos.jpl.nasa.gov/stats/totals.html), this ratio was almost the same (1.74) for a larger number of NEOs. The ratio of the number of ECOs to that of NEOs considered by Granvik et al. (2018), 0.58, yields a reciprocal quantity of 1.72, which coincides with the value for the ratio of the number of NEOs to that of ECOs considered here.

In many studies (e.g., Werner and Ivanov, 2015; Emel'yanenko and Naroenkov, 2015; Granvik et al., 2018), the probability of a collision of an ECO with the Earth was considered to be close to $(3-4) \times 10^{-9}$. According to Morbidelli et al. (2020), the probability that a NEO will impact the Earth in a year is $1.33 \times 10^{-9}$. If $k_{ECO} = 1.72$, this probability for the ECO corresponds to $p_E \approx 2.3 \times 10^{-9}$. These estimates of $p_E$ are 2–5 times smaller than those based on the consideration of orbits of large ECOs observed. The latter were made by Bottke et al. (1994) and Dvorak and Pilat-Lohinger (1999), who used the algorithms close to that of Wetherill (1967), and by Ipatov (2000, 2001) and Ipatov and Mather (2004a, 2004b), who used quite a different algorithm to calculate the probabilities of collisions between bodies and planets. The probability that an ECO will impact the Earth in a year is $p_E = 1/T_E$, where $T_E$ is the characteristic time from the current moment to a collision of the ECO with the Earth. The number of ECOs known even in 2001 (particularly, those considered by Ipatov and Mather (2004a, 2004b)) exceeded the presently known number of ECOs with diameters not smaller than 1 km, while the consideration of more ECOs reduced $T_E$ only slightly. In the papers by Bottke et al. (1994), Dvorak and Pilat-Lohinger (1999), Ipatov (2000, 2001) and Ipatov and Mather (2004a, 2004b), the collision probabilities of the observed ECOs with the Earth were calculated (with known values of the semimajor axes, eccentricities, and orbit inclinations of these ECOs) and the average of this probability for a year $1/T_E$ was determined. The value of $p_E$ obtained in those papers was close to $10^{-8}$.

At the end of this section, we will discuss why there may be no severalfold discrepancy between different estimates of $p_E$; this discussion is based, among other reasons, on the calculations of the collision probabilities of NEOs with the Earth (Emel'yanenko and Naroenkov, 2015). These probability estimates may relate to different time intervals; and the values reported by Werner and Ivanov (2015), Emel'yanenko and Naroenkov (2015), and Granvik et al. (2018) cor-respond to the probabilities that 1-km bodies will fall onto the Earth in the immediate future. For example, Emel'yanenko and Naroenkov (2015) consider a time interval of +/–300 years. These estimates were made when studying the problem of the present asteroid hazard. In this paper, we analyze the craters having been formed over 1 Ga. When the collision probabilities of 1-km bodies with the Earth for millions of years are considered, their distribution over orbit elements may include the elements under which the probabilities of collisions with the Earth are higher than those for the present NEO orbits. For example, according to Emel'yanenko and Naroenkov (2015), who considered the motion of NEOs in an interval of +/–300 years, the gravity focusing effect in the motion of NEOs within some sphere around the Earth was 1.85 times stronger for objects with an absolute magnitude of $H < 27$ than for those with $H < 18$, since the encounters at lower velocities are probable (for the same dynamic characteristics at different values of $H$). As a more comprehensive analysis shows (see below), the present study yields the estimate $T_E \approx 150$ Ma (corresponding to $p_E \approx 6.7 \times 10^{-9}$) for objects with an absolute magnitude of $H < 27$. In our view, though for the recent millions of years the number of 1-km NEOs and their typical orbital elements have remained probably unchanged, the orbital elements of some NEOs could take the values that provided the probabilities of collisions with the Earth and the Moon higher than those based on the current orbits of 1-km NEOs. In the last section of this paper, we discuss how the data on the age of Copernican craters may be interpreted for different values of the collision probabilities of ECOs with the Earth and the Moon for long time intervals (not shorter than millions of years).

As follows from the paper by Granvik et al. (2018, Fig. 26), the probability that asteroids brighter than $18^m$ (i.e., on average, not smaller than 1 km across) will fall onto the Earth is approximately $2 \times 10^{-6}$. In other words, with accounting for the number of ECOs, the probability that a single ECO will impact the Earth is around $4 \times 10^{-9}$. Granvik et al. (2018) estimated the collision probabilities of ECOs with the Earth not for the observed ECOs orbits, but for the orbits of asteroids coming to the Earth's orbit from seven different sources; although they were adjusted to the observed distribution of NEOs orbits. Granvik et al. (2018, Fig. 26) present the curve of the number of collisions with the Earth for the bodies with absolute magnitude smaller than $H$ in dependence on $H$. At $H \approx 25-26$, the points on this curve correspond to the probability values several times (almost an order of magnitude) smaller than those shown in the same figure for the Earth impacts of the Tunguska-type bodies and bolides (according to Brown et al. (2013)). In the paper by Granvik et al. (2018, Fig. 26), the curve of the probability of collisions with the Earth for objects with $17 \leq H \leq 25$ extrapolated to the data for bolides with $H \geq 26$ agrees with the data by Brown et al. (2002), but is inconsistent with the data from a recent paper by Brown et al. (2013). To fit the data by Brown et al. (2013) and the probabilities that Tunguska-type objects will fall onto the Earth, the curve of the collision probability with the Earth for objects with $17 \leq H \leq 25$ should lie much higher than the curve shown by Granvik et al. (2018, Fig. 26). The higher lying curve better agrees with the probability $p_E \approx 10^{-8}$ we use for a collision of an ECO with the Earth over one year (i.e., the probability that an object with $H = 18$ will fall onto the Earth is close to $5 \times 10^{-6}$, since the number of such ECOs is around 500). Harris and D'Abramo (2015) consider the characteristic time period between infalls of Tunguska-type objects onto the Earth to be ~500 years (an event of this kind occurred only about 100 years ago), while according to the curve on Fig. 26 from the paper by Granvik et al. (2018) this characteristic time exceeds 1000 years.

If the curve from the paper by Granvik et al. (2018, Fig. 26) is moved up according to the probability $p_E \approx 10^{-8}$ that an ECO larger than 1 km across will impact the Earth in a year, which was obtained by Bottke et al. (1994), Dvorak and Pilat-Lohinger (1999), Ipatov (2000, 2001) and Ipatov and Mather (2004a, 2004b), (i.e., the characteristic time interval preceding a collision of an ECO with the Earth is assumed to be $T_E \approx 100$ Ma), then the resulting curve will be much more consistent with the current data on the probabilities of bolide infalls (Brown et al., 2013) and the expected frequency of infalls of Tunguska-type objects onto the Earth than the curve in the cited figure.

The dynamic lifetime of ECOs is smaller than $T_E$ due to collisions with other celestial bodies and ejection to hyperbolic orbits. The value of $T_E$ is used below to take the collision probability into account and determine the characteristic time interval preceding a collision of an ECO with the Earth in the model, within which ECOs always remain on their orbits, do not leave them, and collide with no bodies except the Earth.

When calculating $T_E$, Ipatov (1995, 2000, 2001) and Ipatov and Mather (2004a, 2004b) considered all of the ECOs known by the time of simulations. In the paper by Ipatov (1995), the value $T_E = 76$ Ma was obtained for 93 ECOs known in 1991. The value $T_E = 134$ Ma was obtained for ECOs by Bottke et al. (1994), while Ipatov (2000) determined $T_E = 100$ Ma for 417 ECOs. This quantity was estimated as 120 Ma for 54 Apollo asteroids (Dvorak and Pilat-Lohinger, 1999) and 105 Ma for 363 objects of the Apollo group (Ipatov, 2000, 2001). To estimate $T_E$, all of the authors cited above except Ipatov used the algorithms close to that considered by Wetherill (1967), which develops the ideas by Öpik (1951).

Ipatov and Mather (2004a, 2004b) obtained the values $T_E = 15, 164,$ and 67 Ma for 110 Aten asteroids,

643 Apollo asteroids, and all ECOs, respectively. The Amor asteroids are asteroids with perihelion distances larger than 1.017 AU. Consequently, almost all of them cannot approach the Earth (until they change their perihelion distances) to the distance corresponding to the Earth's activity sphere. Temporal variations in the semimajor axes, eccentricities, and orbit inclinations of asteroids were ignored in the calculations. Though the Apollo asteroids accounted for 85% of all ECOs in these calculations, their resulting values of $T_E$ were 2.4 times higher than that for all ECOs. When $T_E = 67$ Ma, the probability $p_E = 1/T_E$ of that ECO will impact the Earth for a year is $1.5 \times 10^{-8}$. The discrepancies between the values of $T_E$ reported in different papers were caused by the differences in the algorithms used and the number of asteroids considered.

The values of $T_E$ obtained by Ipatov and Mather (2004a, 2004b) for all ECOs were smaller than those reported by Ipatov (2000) due to the contribution of several Aten asteroids with small orbital inclinations, which were discovered in the early 2000s. The increase in the orbital inclination of asteroid 2000 SG344 of the Aten group from its current value of 0.1° to 1° resulted in $T_E = 28$ and 97 Ma for the Aten asteroids and all ECOs, respectively (Ipatov and Mather, 2004a). These values of $T_E$ are larger than those obtained for the current orbital inclination of asteroid 2000 SG344, which illustrates the importance of accounting for a small number of asteroids with a high probability of their collision with the Earth (a role of small inclinations is also discussed in the next section). Ipatov (1995, 2000) noted that the value of $T_E$ for all ECOs is several times lower than that calculated for the mean values of inclinations and eccentricities of the same ECOs with the same formulas. For the objects that initially crossed Jupiter's orbit and the resonance asteroids, the analysis of the evolution of their orbits showed that the collision probability for one object of this type with the Earth may be larger than that for thousands of other objects with initially close orbits (Ipatov and Mather, 2003, 2004a, 2004b, 2007).

Ipatov and Mather (2004a, 2004b) calculated $T_E$ for the current orbital elements of ECOs. However, the values analogous to $T_E = 67$ Ma could also be obtained for an interval of 1 Ga, since very likely some ECOs could also have analogously small inclinations before. The orbital inclinations of ECOs vary with time. Ipatov and Mather (2004a, 2004b) considered 110 Aten asteroids. If the orbital inclinations are randomly distributed in a range of 0° to 11°, one of them will be smaller than 0.1° (in the previous paragraph, we discussed the contribution to $T_E$ made by asteroid 2000 SG344 with an orbital inclination of 0.1°). If the inclinations are randomly distributed in a range of 0° to 22°, one of them will be in a range of 0° to 0.2°, i.e., its average will be 0.1°.

When calculating the probabilities $p_{dts}$ of that two celestial objects (e.g., an ECO and the Earth) approach each other to the distance corresponding to the radius $r_s$ of the considered sphere (usually, it is the sphere of action of a planet with a mass $m_{pl}$ and a radius $r_s \approx R(m_{pl}/M_S)^{2/5}$, where $M_S$ is the mass of the Sun) for the time interval $d_t$, the following formulas were used in the three-dimensional model (Ipatov, 1988b; 2000, Ch. 4, Sect. 2): $p_{dts} = d_t/T_3$, where $T_3 = 2\pi^2 \cdot k_p \cdot T_s \cdot k_v \cdot \Delta i \cdot R^2/(r_s^2 \cdot k_{fi})$ is the characteristic time elapsed before the encounter, $\Delta i$ is the angle between the orbital planes of the approaching celestial objects, $R$ is the distance from the encounter point of celestial objects to the Sun, $k_{fi}$ is the sum of the angles (expressed in radians) subtended at the Sun, within which the distance between the projections of orbits (along the line originating in the Sun) is smaller than $r_s$ (this sum varies for different orbits (see, Ipatov, 2000, Fig. 4.1), $T_s$ is the synodic period, $k_p = P_2/P_1$, $P_2 > P_1$, $P_i$ is the rotation period of $i$th object around the Sun, $k_v = (2a/R - 1)^{1/2}$, and $a$ is the semimajor axis of the object's orbit. The coefficient $k_v$ was added by Ipatov and Mather (2004a, 2004b) to account for the dependence of the approach velocity on the object position on the eccentric orbit. The collision probability for the objects that entered the sphere of action was assumed to be $p_{dtc} = (r_\Sigma/r_s)^2(1 + (v_{par}/v_{rel})^2)$, where $v_{par} = (2Gm_\Sigma/r_\Sigma)^{1/2}$ is the parabolic velocity, $v_{rel}$ is the relative velocity of objects that approached each other to the distance $r_s$, $r_\Sigma$ is the sum of the radii of colliding objects with the total mass $m_\Sigma$, and $G$ is the gravitational constant. For small values of $\Delta i$, the other formulas were used in the algorithm. The algorithms (and their basis) to calculate $k_{fi}$ and the characteristic time between the collisions of objects were presented by S.I. Ipatov in Report O-1211 of the Keldysh Institute of Applied Mathematics (1985, Appendix 3, p. 86–130). The probability that an ECO will impact the Earth over the time interval $d_t$ is $p_{dt} = p_{dts} \times p_{dtc}$. The values of $p_{dts} = d_t/T_3$ and $T_3$ were calculated for the fixed values of semimajor axes, eccentricities, and orbital inclinations of the considered celestial objects, but for various possible orientations of their orbits. When calculating $p_E = 1/T_E$, the $p_{dt}$ value averaged over all considered ECOs was used.

The algorithms for calculating the probabilities of collisions of bodies with planets were used by Ipatov (2000, 2019b) to study the process of accumulation of planets. These algorithms are oriented to considering the collision probabilities of numerous planetesimals over long time periods. Averaging over various orientations of orbits (under fixed values of the semimajor axes, eccentricities, and orbital inclinations) was included into these algorithms. The estimates of the time for growth of planets agree well with the data of different authors. For example, from calculations of these collision probabilities, Ipatov (2019b) inferred that half the mass of the Earth and Venus was accumulated over a period not larger than 5 Ma, while Lammer et al. (2019) independently came to the same conclusion from the analysis of the isotope composition of noble gases in the atmospheres of these planets. Our estimates of $p_E = 1/T_E$ are consistent with numerous models of the Earth's accumulation. In most calculations of the Earth's accumulation from planetesimals (both for analytical estimates and in numerical integration of the equations of motion or accounting for the gravitational interaction with the method of spheres) performed by different authors (see, Ipatov (2000, 2019) and references therein), almost all surrounding bodies fell onto the Earth over a time not exceeding 100 Ma. For the past 60 years, this value (100 Ma) has been generally accepted for the time of almost complete accumulation of the Earth. The model results for the evolution of disks of the gravitating bodies, which are coagulated in collisions, showed that, in the final stages of the Earth's accumulation, the mean eccentricities of orbits of planetesimals could exceed 0.4 (Ipatov, 1993), i.e., they might be comparable to the orbital eccentricities of present ECOs. From the velocities of collisions of planetesimals with the almost completely formed Earth, Ipatov (2019b) concluded that the orbital eccentricities of most planetesimals that impacted the Earth exceeded 0.3. Because of this, it can hardly be supposed that the values of $T_E$ for planetesimals, which fell onto the almost completely formed Earth, were substantially smaller than those for presently existing ECOs.

The above discussion, as well as the comparison made below for the collision probabilities of asteroids with the terrestrial planets obtained with the algorithm by Wetherill (1967), which was used by Bottke et al. (1994) and Dvorak and Pilat-Lohinger (1999), and the numerical integration of the equations of motion allow us to assume the following. To estimate the probability that an ECO larger than 1 km across will impact the Earth in a year, we may use, as one of the probable values, $p_E \approx 10^{-8}$ together with that close to $(3–4) \times 10^{-9}$ (i.e., we may consider the characteristic time before a collision as $T_E \approx 100$ Ma).

In many papers, to calculate the collision probability of bodies with a planet, the approach based on the ideas by Öpik, specifically, the modification of the Öpik equation made by Wetherill (1967), was used. This modification with some corrections was considered in several subsequent papers. Dones et al. (1999) compared the collision probabilities of the migrated bodies with planets, which were found with the algorithm based on that by Wetherill (1967), to the results of calculations with numerical integration. In particular, the probabilities that the bodies ejected from the Moon will collide with Venus and the bodies ejected from Mars will collide with the Earth turned out to be somewhat higher according to the numerical integration than those from applying the Öpik method (Dones et al., 1999, Fig. 6). Vokrouhlický et al. (2012) and Pokorny and Vokrouhlický (2013) considered the modified algorithm of Wetherill (1967), which took into account the changes in the argument of perihelion of an asteroid due to the influence of Jupiter and other planets. We note that our algorithm also used the value averaged over different values of the argument of perihelion. Pokorny and Vokrouhlický (2013) compared the numbers of collisions of bodies with the terrestrial planets calculated with three methods: the numerical integration of the equations of motion, the algorithm of Wetherill (1967), and the abovementioned modification of that algorithm. Based on the data from Table 1 from (Pokorny and Vokrouhlický, 2013) in Table 4, we present the ratios of the number of collisions of the e-belt asteroids (asteroids with large semimajor axes of orbits ranging from 1.7 to 2.1 AU) with planets for calculations with the algorithm of Wetherill (1967) and its modification to the number of collisions obtained by numerical integration. The data from Table 4 show that the modified algorithm yields a little more accurate values for the collision probabilities than the algorithm of Wetherill (1967). The latter overestimated the collision probability by 8% for the Earth and underestimated it roughly by 1.5 times for Mercury and Mars. The larger errors for Mercury and Mars are apparently connected with large eccentricities and inclinations of orbits of these planets, for which the Öpik method is less accurate. The results of Dones et al. (1999) and Pokorny and Vokrouhlický (2013) allow one to suggest that the characteristic time elapsed before a collision of an ECO with the Earth estimated as being close to 100 Ma by the algorithm analogous to that of Wetherill (1967) (i.e., the estimates of the probability of a collision of an ECO with the Earth for a year are close to $10^{-8}$) will hardly differ by more than 1/3 from the estimates obtained in numerical integration. Werner and Ivanov (2015) considered $T_E \approx 300$ Ma. We have failed to understand why Bottke et al. (1994), Dvorak and Pilat-Lohinger (1999), and Werner and Ivanov (2015) obtained the values of $T_E$ differing by almost three times, though they referred to the same algorithm of Wetherill (1967). The fact that they considered different sets of ECOs could hardly result in such large discrepancies.

As distinct from the abovementioned studies, which used some formulas to estimate the probabilities of collisions of bodies with planets, Emel'yanenko and Naroenkov (2015) estimated the collision probability of NEOs with the Earth from the evolution of orbits of known NEOs numerically integrated over an interval of 600 years. The number of approaches of NEOs to the Earth to a distance of 0.05 AU and the relative velocities in these approaches were calculated. In the paper by Emel'yanenko and Naroenkov (2015), for the objects with absolute magnitudes higher than $H = 18$, the frequency of collisions with the Earth was obtained as 0.53 Ma. If the currently estimated number of 1-km NEOs is 920 and the ratio between the numbers of

NEOs and ECOs is 1.72, a frequency of 0.53 Ma corresponds to $T_E$ = 283 Ma. In our view, the results of calculations by Emel'yanenko and Naroenkov (2015) allow the values of $T_E$ to be smaller (and the values of $p_E$ to be larger), if the probability of collisions between 1-km NEOs and the Earth is considered on time intervals much larger than 600 years. It is our opinion that the distribution of 1-km NEOs over their orbits for millions of years could be close to the distribution of numerous small objects for hundreds of years, and a large contribution to the total probability of collisions of NEOs with the Earth (or the Moon) could be made by a small number of NEOs in those time intervals when they moved along the orbits characterized by relatively large collision probabilities with the Earth (or the Moon), for example, along the orbits lying almost in the ecliptic plane.

In the paper by Emel'yanenko and Naroenkov (2015), for objects with $H < 27$, the coefficient accounting for the gravity focusing due to the gravitational field of the Earth (as compared to the case ignoring this factor) was 1.846 times larger (i.e., 240/130) than that for objects with $H < 18$, since NEOs may approach the Earth with lower velocities. Under this gravity focusing, $T_E$ for 1-km ECOs would be 150 Ma (i.e., 283/1.846) instead of 283 Ma. If we consider a time interval of hundreds of millions of years instead of 600 years, we may assume that small relative velocities will appear in the distribution over approach velocities of 1-km ECOs with the Earth, as in the case of larger statistics of approaches of smaller NEOs to the Earth for 600 years. However, when considering 1-km NEOs over long time intervals, the difference in the gravity focusing may be smaller than that in the comparison to the objects with $H < 27$, since it is easier to discover the slowly moving objects with $H < 27$. When accounting for the gravity focusing at distances from the Earth not exceeding several Earth radii, Emel'yanenko and Naroenkov (2015), as many other authors, used a formula for the hyperbolic motion of bodies near the Earth. If the relative velocities are low, the relative motion may be more complex rather than hyperbolic; and the body may even make several revolutions around the planet. Potentially, the collision probability under small relative velocities may be larger than that for the trajectory of the relative motion represented by a hyperbola. Large statistics of collisions may substantially reduce the characteristic time $T_E$ elapsed before collisions of bodies with planets. Moreover, what is important is the increase in the portion of approaches of NEOs to the Earth or the Moon with low relative velocities rather than the change in the mean value of these velocities. Under the same values of the mean approach velocities, the mean value of the coefficient accounting for the gravity focusing may be larger if the minimal approach velocities are smaller or/and the portion of bodies with minimal approach velocities is larger. In addition, the values of $T_E$ may be smaller if the inclinations of heliocentric orbits of NEOs are smaller.

Table 4. The ratios of the number of collisions of the e-belt asteroids with the planets $N_{Weth}$ calculated with the algorithm of Wetherill (1967) and the corresponding number obtained with the modification of this algorithm $N_{PV}$ to the analogous number of collisions obtained by numerical integration $N_{direct}$ (according to the data by Pokorny and Vokrouhlický (2013))

| Planet | Mercury | Venus | Earth | Mars |
|---|---|---|---|---|
| $N_{Weth}/N_{direct}$ | 0.68 | 1.00 | 1.08 | 0.67 |
| $N_{PV}/N_{direct}$ | 0.91 | 1.00 | 1.06 | 1.08 |

As was noted by Ipatov and Mather (2003, 2004a, 2004b, 2006) and Ipatov (2019a), the main contribution to the value of the collision probability of bodies with the Earth during the dynamical lifetimes of bodies can be made by a small portion of the considered migrating bodies, and the contribution of a single body to the total probability of collisions of bodies with the Earth may be even higher than that for hundreds or thousands of other bodies with close initial orbits. For example, in the paper by Ipatov (2019a), the probabilities of collisions $p_E$ of planetesimals from the feeding zone of Jupiter and Saturn with the Earth for the dynamical lifetimes of planetesimals could differ by dozens of times for different groups containing 250 planetesimals with close initial orbits. The smaller the time interval, the larger may be the relative contribution to $p_E$ from a small number of bodies. As has been mentioned above, according to the data by Emel'yanenko and Naroenkov (2015), the gravity focusing coefficient for objects with $H < 27$ was 1.846 times higher than that for $H < 18$. It was also assumed that the dynamical characteristics of smaller NEOs are the same as those for objects with $H < 18$. In our opinion, the possibilities that individual ECOs may transfer to the orbits where the collision probability with the Earth (or the Moon) was higher than the total probability for hundreds of other ECOs may contribute more (relative to the increase in the gravity focusing coefficient) to the possibility of increasing the probabilities of collisions of 1-km NEOs with the Earth (or the Moon) in the consideration of an interval of millions of years as compared to the estimates of the probability for hundreds of years.

In calculations for NEOs with $H < 27$, Emel'yanenko and Naroenkov (2015) considered a value of ~8 km/s for the mean velocity of approaching the Earth to a distance of 0.05 AU (it was also noted that this observed value is substantially influenced by the observational selection). For a parabolic trajectory, this mean velocity corresponds to a value of ~13.8 km/s for the velocity of impacting the Earth. Drolshagen et al. (2020) noticed that, in the papers they cited, the peak of the distribution of velocities of bodies impacting the

Earth was at ~13–15 km/s, while the velocities of meteorites entering the terrestrial atmosphere may change in a range of 11 to 73 km/s, and meteoroids with velocities of 9.8 and 10.9 km/s were also detected. This means that some meteoroids approach the activity sphere of the Earth with almost zero velocity, since the parabolic velocity on the Earth's surface is 11.2 km/s.

A velocity of 13.8 km/s is smaller than the collision velocities considered above, 19 km/s, which corresponds 15-km craters produced by 1-km impactors in our calculations. For an impactor 1 km in size, a velocity of 13.8 km/s corresponds to the 13-km diameter of a lunar crater. Table 1 contains only two craters larger than 12 and smaller than 15 km across for the entire lunar surface. In Table 2, four craters on maria exhibit diameters between 12 and 17 km. Consequently, the number of craters with diameters $D \geq 13$ km does not differ much from that with $D \geq 15$ km. The mean sizes of craters also depend on the velocity distribution of impactors rather than only the mean values of collision velocities.

The ratio of the probabilities of collisions of NEOs with the Earth to those of NEOs with the Moon is believed to be $p_{EM} \approx 22$ (Le Feuvre and Wieczorek, 2011). Close estimates of $p_{EM}$ were obtained by Ipatov (2019b).

## COMPARISON OF THE NUMBER OF OBSERVED LUNAR CRATERS TO THAT CALCULATED FROM THE NUMBER OF NEAR-EARTH OBJECTS AND THE PROBABILITIES OF THEIR COLLISIONS WITH THE MOON

### The Number of Craters Estimated from the Data on Near-Earth Objects

The number of infalls of NEOs with diameters $d > 1$ km onto the considered lunar area over the time $T$ can be estimated with the formula

$$N_{est} = N_{1av} r_{reg} T / (T_E p_{EM} k_{ECO}), \qquad (6)$$

where $N_{1av}$ is the mean number of NEOs with diameters $d > 1$ km for the time $T$, $k_{ECO}$ is a portion of ECOs among NEOs, $p_{EM}$ is the ratio of the collision probabilities of ECOs with the Earth and the Moon, $p_E = 1/T_E$ is the probability that an ECO will impact the Earth over a year, and $r_{reg}$ is the ratio of the considered area to that of the Moon. Formula (6) was obtained for the model, within which the number of ECOs and the probabilities of their collisions with the Moon are constant during the considered time interval $T$. For $N_{1av} = N_1 = 920$, $k_{ECO} = 1.72$, $T = 1100$ Ma, $T_E = 100$ Ma, and $p_{EM} = 22$, we obtain $N_{est} = N_{est-sea} = 920 \times 0.155 \times 1.1/(0.1 \times 22 \times 1.72) \approx 41.45$ if $r_{reg} = 0.155$ and $N_{est} = N_{est-ful} \approx 267$ for $r_{reg} = 1$. For these data, the number of 1-km ECOs is assumed to be $920/1.72 = 535$. Analogously, for $T_E \approx 300$ Ma, we have $N_{est-ful} \approx 89$ and $N_{est-sea} = 14$. The value $T_E \approx 300$ Ma corresponds to a probability of $(3-4) \times 10^{-9}$ that an ECO with $H < 18$ will fall onto the Earth during a year (Werner and Ivanov, 2015). If $T_E = 67$ Ma, then $N_{est-ful} \approx 398.5$ and $N_{est-sea} \approx 61.9$.

### The Number of Copernican Craters on Maria and Highlands

If we proceed from the data by Mazrouei et al. (2019a) and Losiak et al. (2015), which are related to the entire surface of the Moon, the estimates of the number of craters not smaller than 15 km across and younger than 1.1 Ga are $N_{obs} = 44$ and 53, i.e., six and five times smaller than 267 (this is the $N_{est-ful}$ estimate at $T_E = 100$ Ma), respectively. For $T_E = 300$ Ma, $N_{est-ful}$ will exceed $N_{obs}$ approximately twofold. The difference between $N_{obs}$ and $N_{est-ful}$ is smaller, if the estimates of the crater number are considered under the assumption that the crater number density is the same as that in the Oceanus Procellarum and other maria on the near side of the Moon. The estimates of this kind for the number of craters not larger than 15 and 18 km in diameter and younger than 1.1 Ga correspond to the values of $N_{reg15}/r_{reg}$ and $N_{reg18}/r_{reg}$ from Table 3. The values of $N_{reg15}/r_{reg}$ and $N_{reg18}/r_{reg}$ derived from the analysis of the craters in the Oceanus Procellarum and other maria on the near side of the Moon considered by Mazrouei et al. (2019a) do not exceed 77. These values are smaller than 267 (this is the $N_{est-ful}$ estimate at $T_E = 100$ Ma), 398.5 (at $T_E = 67$ Ma), and 89 (at $T_E = 300$ Ma) at least by 3.5, 5.2, and 1.16 times, respectively. According to Losiak et al. (2015), 53 lunar craters with diameters not smaller than 15 km are dated as Copernican; and 29 of them (more than half) are in the mare region on the near side of the Moon. If the crater density is the same, these 29 craters in the mare region represent 187 craters on the entire lunar surface. Even this maximal estimate (187) of the crater number is smaller than 267 (the estimate of $N_{est-ful}$ at $T_E = 100$ Ma) and 398.5 (at $T_E = 67$ Ma) by 1.4 and 2.1 times, respectively, but 2.1 times larger than 89 (at $T_E = 300$ Ma).

As has been noted above, the number density of Copernican craters on the maria are higher than that on the highlands. Apparently, this difference is partially connected with the inaccuracy in determining the age of craters, especially outside the lunar maria; and the number density corresponding to the entire lunar surface may be larger than that according to Mazrouei et al. (2019a) and Losiak et al. (2015). Since the differences in the values of the number density for maria and highlands were obtained by different authors, these values may actually differ. For example, the differences in the processes of formation (a larger diameter of a transient crater) and modification of craters in mare and highland regions (collapsing craters on maria become wider than those on highlands for 1 Ga) may be caused by differences in the characteristics of the underlying surface where craters are formed. Lunar maria are flat-floor lowlands filled with solidified lava. Highlands are covered with volcanic rocks—basalts.

Different widening of collapsing craters on maria and highlands may be characterized by a smaller depth of mare craters. With formulas from the paper by Kalynn et al. (2013), we estimate the depths of fresh craters with $D = 15$ km on maria and highlands as 2.26 and 3.1 km, respectively. According to the data from Table 1, the depths of mare and highland craters with diameters close to 15 km and ages smaller than 1.1 Ga are in ranges of 2.2–3 and 2–2.4 km, respectively. In Table 2, the depths of mare craters of this kind range from 2.2 to 2.5 km. In other words, according to the data from Tables 1 and 2, Copernican craters with diameters close to 15 km on maria and highlands exhibit no sharp difference in depths. As distinct from the paper Kalynn et al. (2013), Tables 1 and 2 contain the data for Copernican craters. Kalynn et al. (2013) selected 111 craters with diameters from 15 to 167 km among the craters younger than 3.2 Ma according to at least one of the following criteria they proposed for the crater structures: (1) impact melt on the crater floor and ejecta facies are observed; (2) a crisp crater rim is well defined; (3) the crater walls exhibit distinct faults; and/or (4) rays are seen in the ejecta blanket. Kalynn et al. (2013) believed that their criteria make it possible to select the freshest craters.

In this paper, we do not analyze the diameter distribution of all lunar craters (i.e., the Neukum curve as a whole) but consider only the number of Copernican craters with diameters larger than 15 km on the basis of our calculations. For the other sizes and ages of craters, the ratio of the number of observed craters to that estimated from the present number of ECOs and the probabilities of their collisions with the Moon may differ from the ratio considered here.

*Estimates of the Accuracy in Determining the Number of Craters for the Considered Model*

Main errors in determining the crater number with formulas (5) and (6) are connected with the inaccuracy in calculations of the collision probabilities of NEOs with the Moon. As has been mentioned above, some authors assume a value of $p_E \approx (2–4) \times 10^{-9}$ for a collision of one ECO with the Earth over a year, while the others believe that $p_E$ is close to $10^{-8}$ (with deviations of 1.5 times up or down). Consequently, different authors considered the values of $T_E = 1/p_E$ to be close to 300 or 100 Ma. In our opinion, larger estimates of $p_E$ may appear in dealing with large time intervals. The both values of $T_E$ are used below to compare the results of model calculations with the data on the ages of lunar craters. Apparently, the real values of $T_E$ and $p_E$ are in the interval between those specified above. Some of the conclusions made below are valid for the both values of $T_E$. Even at a fixed number of NEOs, the values of $T_E$ could vary with time due to changes in the distribution of orbital elements of NEOs.

In our view, the other inaccuracies connected with determining the number of craters with formulas (5) and (6) are substantially smaller than the uncertainty in estimating $T_E$. The errors in determining the dependence of the crater diameter on the impactor diameter probably did not exceed 5–10%. Depending on the albedo, the diameters of NEOs with the absolute magnitude $H = 17.75$ may differ by several times. The number of the currently existing NEOs, which are not smaller than 1 km across, was estimated under the mean albedo value, 0.14. In the book *Asteroidno-Kometnaya Opastnost'* (2010, p. 68), it is noted that, for individual asteroids, the smallest and highest values of the albedo may differ from the mean one roughly by five times. Moreover, the limiting values of diameters may differ by ~2.25 times from the nominal one corresponding to the mean albedo.

For the bodies with $H = 17.75$, which are fragments of the catastrophic destruction of an asteroid, the mass may change by four times depending on the type (S or C), while the crater diameter for a velocity of 20 km/s and $H = 17.75$ ranges from ~15 to ~18 km according to formula (5). If the Neukum curve slope is about –1.25 at $D \approx 16–18$ km, the number of craters will change by ~25%. According to Table 3, the numbers of craters with diameters larger than 15 and 18 km differ by not much than 1.4 times. If the scattering in the densities is accounted for, the difference will be even higher and may reach 1.5 times. This case would occur if all of the craters were formed in several catastrophic collisions of the bodies belonging to a certain type, since the densities and sizes of several parent bodies may differ from the mean values. In our estimates of $N_{est}$, the analogous coefficient of uncertainty may be lower, since $N_{est}$ was estimated for a time interval of 1.1 Ga, during which collisions of many asteroids with various compositions occurred. It can hardly be admitted that the mean composition of NEOs has cardinally differed for the recent 1 Ga.

In dependence on the contribution of destroyed parent asteroids of various composition to NEOs, the distribution of NEOs over albedo could change. If for the recent billion years NEOs have been mostly represented by objects, the albedo of which exceeds 0.14, the present number of NEOs with $H < 17.75$ would yield the somewhat lower number of craters with $D > 15$ km than that based on our estimates. However, the influence of variations in the characteristic albedo and the type of NEOs on the estimate of the number of craters would be noticeable only in the consideration of a time interval much smaller than 1 Ga. To estimate the number of produced craters more accurately, it is better to use the distribution over collision velocities rather than their mean values. Nevertheless, since the $T_E$ value is not accurately determined in our analysis, this consideration will hardly improve general estimates.

For craters with $D > 15$ km formed during 1.1 Ga over the entire lunar surface, the ratio of $N_{est}$ to the number of craters obtained from the papers by Losiak et al. (2015) and Mazrouei et al. (2019a) $N_{est}/N_{obs}$ is close to 5–6 and 2 for $T_E = 100$ and 300 Ma, respectively. Because of this, in spite of different above-listed inaccuracies in determining the number of craters for the present model, we may state that, most likely, the ratio $N_{est}/N_{obs}$ is not smaller than two. This possibility was admitted by Ipatov et al. (2020) when considering the craters with the first integrity degree from the SAI Morphology Catalog of Lunar Craters. According to the "new chronology" (Mazrouei et al., 2019a), the number of Copernican craters roughly halved as compared to that in the paper by Werner and Ivanov (2015). If we accept that the number of Copernican craters younger than 1 Ga in the new chronology is true, the crater formation rate in the "standard" chronology by Neukom should be reduced approximately by one half.

### The Possibility of Increasing the Crater Formation Rate for the Recent 290 Ma

The main result of the paper by Mazrouei et al. (2019a) is that the frequency of collisions of near-Earth asteroids with the Moon increased by 2.6 times 290 Ma ago. For the model, within which the collision probability of a NEO with the Moon for the recent 290 Ma has been equal to the current value while it had been 2.6 times less for 810 Ma before, the number of craters produced would account for 0.6 of the estimate derived from the present number of NEOs (i.e., would be 1.7 times smaller). If the number density of craters is equal to that for the maria according to the data by Losiak et al. (2015) or Mazrouei et al. (2019a), then the number of 15-km craters having been produced over 1 Ga, which is 1.7 times less that the present number of craters, corresponds to $T_E = 100 \times (267/187)/1.7 = 82$ Ma or $T_E = 100 \times (267/77)/1.7 = 200$ Ma for data of these papers. If the number density of craters is equal to that for the whole lunar surface according to the data by Mazrouei et al. (2019a) or Losiak et al. (2015), the number of craters having been produced over 1 Ga, which is 1.7 times less that the present number of craters, corresponds to $T_E = 100 \times (267/44)/1.7 = 357$ Ma or $T_E = 100 \times (267/53)/1.7 = 296$ Ma, respectively. The last specified estimates are close to $T_E = 300$ Ma (which approximately corresponds to that considered by Werner and Ivanov (2015) and Emel'yanenko and Naroenkov (2015)). Thus, if $T_E \le 300$ Ma, all of the above analyzed estimates of $N_{est}$ and $N_{obs}$ allow the probability of a collision with the Moon to grow by 2.6 times 290 Ma ago. Our estimates at $T_E \le 100$ Ma agree better with this conclusion by Mazrouei et al. (2019a), if the number density of craters on the entire lunar surface is assumed to be the same as that for craters in the region of the Oceanus Procellarum and the maria on the near side of the Moon. Consequently, we may suppose that the crater number density for the entire lunar surface could be almost the same as that for the abovementioned region, i.e., could be larger than the corresponding estimates obtained from the data by Losiak et al. (2015) or Mazrouei et al. (2019a) for the entire lunar surface.

The results of comparison of the crater formation models to the crater age estimates leave open the possibility that the frequency of collisions of near-Earth asteroids with the Moon increased by 2.6 times 290 Ma ago and, moreover, does not contradict the growth of the collision frequency for a different time interval (not necessarily 290 Ma). Though our estimates do not contradict the conclusion by Mazrouei et al. (2019a) that the frequency of collisions of near-Earth asteroids with the Moon increased by 2.6 times 290 Ma ago, they do not prove this conclusion, since they admit that, over a whole period of 1.1 Ga, the real number of Copernican lunar craters formed per time unit could be several times larger than that according to the estimates by Mazrouei et al. (2019a). These authors explain the 2.6-fold increase in the frequency of collisions of near-Earth asteroids with the Moon occurred 290 Ma ago by catastrophic destruction of large main-belt asteroids that may have occurred over the recent 300 Ma.

According to Bottke et al. (2007), the recent catastrophic destruction of a large main-belt asteroid occurred 160 Ma ago could enhance the present number of NEOs with diameters $d > 1$ km as compared to the mean number of NEOs with diameters $d > 1$ km obtained for the 1-Ga interval. Our estimates do not contradict these statements, because, according to these estimates, the number of observed craters $N_{obs}$ is smaller than the number of craters $N_{est}$ obtained from the present number of NEOs. However, a majority of bodies formed in fragmentation of the asteroid 160 Ma ago might no longer be on the NEOs orbits. Over this time they have either transferred to hyperbolic orbits, or remained in the asteroid belt, or collided with other celestial bodies.

Gladman et al. (2000) obtained that the median lifetime of NEOs is around 100 Ma. In the paper by Ipatov (2019b), for the bodies with semimajor axes of initial orbits not exceeding 1.5 AU and initial eccentricities of 0.05 and 0.3, the median dynamical lifetime did not exceed 20 Ma. Consequently, the destruction of the asteroid, which induced the present increase in the number of NEOs (in relation to the mean value for the recent billion years), could occur relatively recently rather than 160 Ma ago, as Bottke et al. (2006) believed. To provide the enhancement in the current number of NEOs due to the asteroid destruction that occurred 160 Ma ago, it should be assumed that most fragments formed in this destruction started to cross the Earth's orbit more than 100 Ma after the destruction. Bottke et al. (2006) supposed that the fragments migrated to the 7 : 2 and 5 : 9 resonances with Jupiter and Mars, respectively, due to the Yarkovsky and Yarkovsky–O'Keefe–Radzievskii–Paddack (YORP) effects. Mazrouei et al. (2019a) also assumed that the migration of fragments to the resonances, which transferred the fragments to the Earth, was slow because of nongravitational forces.

Nesvorny et al. (2002) believed that the family of asteroid (832) Karin was formed 5.8 Ma ago. The semimajor axis, eccentricity, and inclination of the orbit of this asteroid are 2.865 AU, 0.08, and 1°, respectively. This family is not far from the 5 : 2 resonance with Jupiter (2.82 AU). Many bodies captured into this resonance relatively quickly increase their eccentricities and reach the orbits of Mars and the Earth (Ipatov, 1988a, 1989, 1992a, 1992b, 2000; Morbidelli and Gladman, 1998). Zappala et al. (1998) estimated the time for which fragments of asteroid families transfer to resonance orbits in a range from 0.3 to 110 Ma for Gelion, Dora, and Koronis families (the 5 : 2 resonance) and the Eos family (the 9 : 4 resonance), respectively; their estimate of the "asteroid shower" induced by the fragments ranges from 2 to 30 Ma. According to these authors, 90% of bodies left the 5 : 2 and 3 : 1 resonances with Jupiter over 5 and 10 Ma, respectively. Migliorini et al. (1997) found that the median lifetimes of the Vesta family fragments, which possessed orbits corresponding to the $v_6$ and 3 : 1 resonances, are 2 Ma. According to Ipatov and Mather (2004a), the times, during which the aphelian distances of asteroids were smaller than 4.2 AU, did not exceed 2 and 9 Ma for asteroids that had started from the 5 : 2 and 3 : 1 resonances with Jupiter, respectively, while the mean lifetime of objects, which had crossed Jupiter's orbit, were about 0.1 Ma. Milani and Farinella (1995) came to conclusion that the bodies, being beyond the 5 : 2 resonance with Jupiter at a distance of 0.001 AU, may be trapped in the resonance for several million years due to chaotic diffusion. Based on the results described above in this paragraph, we suppose that the destruction of a small body, which increased the number of observed NEAs, could occur not earlier than 10–20 Ma ago. For example, this destruction could take place 6 Ma ago, when the Karin family was formed. The above estimates of the migration time of bodies from the asteroid belt to the Earth's orbit and the lifetime of NEOs do not usually exceed 10–30 Ma, which is substantially less than 290 Ma considered by Mazrouei et al. (2019a).

Vokrouhlický et al. (2017) supposed that roughly 1 Ga ago the Vesta asteroid family was formed and the influx of fragments to the Earth was highest for the first 100–300 Ma after the family formation. However, in this case, the present influx of NEOs would be provided by quite different sources. For the model by Vokrouhlický et al. (2017), the cratering peak on the Moon would occur much earlier than in the recent 290 Ma, as considered by Mazrouei et al. (2019a). To achieve agreement with the results by Mazrouei et al. (2019a), it is necessary that the influx of fragments to the Moon increased within the last 290 Ma and not in a shorter time interval.

If there were a collision between asteroids with the densities higher than the mean density of asteroids, then somewhat larger than usual craters would be produced on the Moon by their fragments for some time after the collision. However, it seems highly unlikely that a single collision of asteroids was a source of so long-lasting (300 Ma) an enhancement in the number of new lunar craters. It is also difficult to explain why the mean number of mutual collisions of the main-belt asteroids have been doubled for the recent 300 Ma as compared to the averages of the mean number for the previous 700 Ma. Apparently, the question arises whether the current estimates of the age of craters, especially outside the maria, should be corrected. As has been noted in the Introduction, Fig. 1 shows that the data by Losiak et al. (2015) underestimate the number of craters smaller than 30 km across. However, for the entire lunar surface, especially for the maria, the data by Mazrouei et al. (2019a) yield even a smaller number of craters 15 km across than the data by Losiak et al. (2015).

The bombardment spike corresponds to a higher-lying Neukum curve. Moreover, in dependence of the composition, mass, and orbits of asteroids that induced this spike, the shape of this curve may change somewhat with time. This is connected with the mass and composition distributions of fragments at different collisions, and this curve also depends on the time during which these fragments reach the orbits of NEOs. After the fragments originating in different regions of the asteroid belt have become NEOs, their distributions over orbital elements may be different. This distribution may produce some effect on the probability and velocity of a collision of an NEO with the Moon.

It is seen from Fig. 1 that our estimates of the number density of Copernican craters with diameters not smaller than 15 km at $T_E$ = 100 Ma agree rather well with the curve, which would continue the curve for the data for mare regions by Losiak et al. (2015) from the values at $D$ > 30 km to those at smaller $D$ in parallel to the curve by Werner and Ivanov (2015). In other words, our estimates at $T_E$ = 100 Ma correspond to the model, according to which the number density of Copernican craters for the whole lunar surface would be the same as that for maria regions if the data by Losiak et al. (2015) for $D$ < 15 km were as complete as those for $D$ > 30 km. For this model, the cratering rate may have been constant over the recent 1.1 Ga to agree with our calculations of the cratering at $T_E$ = 100 Ma. In this case, there is no need to invent explanations of a larger cratering rate for the recent 290 Ma than that for the previous 800 Ma.

As has been noted in the Section *Probabilities of Collisions of Near-Earth Objects with the Moon*, if the curve for the dependence of the annual number of collisions of bodies with the absolute magnitude smaller than $H$ with the Earth on $H$ (Granvik et al., 2018, Fig. 26) is moved up according to the probability $p_E$ that an ECO larger than 1 km across will impact the Earth during a year, which is $10^{-8}$ (i.e., at $T_E = 100$ Ma), then the curve will agree much better with the present data on the probabilities of bolide infalls onto the Earth (Brown et al., 2013) and the expected frequency of infalls of Tunguska-like objects than the curve shown in that figure.

## CONCLUSIONS

We compared the number of lunar craters larger than 15 km across and younger than 1.1 Ga to the estimates of the number of craters that could have been formed over that time, if the number of NEOs larger than 1 km across and their orbital elements during that time were close to their current values. The comparison was made for craters on the entire lunar surface and in the region of the Oceanus Procellarum and the maria on the near side of the Moon. To make these estimates, we used the values of the collision probabilities of NEOs with the Moon and the dependences of the crater diameters on the sizes of impactors that produced these craters.

It has been noticed that, according to the data by different authors, the number density of known Copernican craters with diameters $D \geq 15$ km on maria is at least the double the corresponding number for the entire lunar surface.

Our estimates do not contradict the enhancement in the number of NEOs after probable catastrophic destruction of large asteroids in the main belt, which may have been occurred over the past 300 Ma; however, our estimates do not prove this enhancement. In particular, they do not contradict the inference by Mazrouei et al. (2019a) that the frequency of collisions of near-Earth asteroids with the Moon increased by 2.6 times 290 Ma ago.

The number of Copernican lunar craters with diameters no smaller than 15 km may be larger than that according to the data Mazrouei et al. (2019a).

If the probability that an ECO will impact the Earth in a year is $10^{-8}$ (this probability might be true for the consideration of large time intervals), our estimates of the crater number correspond to the model within which the number density of 15-km Copernican craters for the whole lunar surface would be the same as that for the mare region if the data of Losiak et al. (2015) for $D < 30$ km were as complete as those for $D > 30$ km. Under this probability of a collision of an ECO with the Earth and for this model, the cratering rate may have been constant over the recent 1.1 Ga.

ACKNOWLEDGMENTS

The authors are grateful to reviewers for numerous valuable comments that contributed much to improving the paper.

FUNDING

The study was performed under government contracts of the Vernadsky Institute of Geochemistry and Analytical Chemistry of the Russian Academy of Sciences, the Sternberg Astronomical Institute of Moscow State University, and the Institute of Geosphere Dynamics of the Russian Academy of Sciences. The studies of the probabilities of asteroid collisions with the Earth were supported by the grant 075-15-2020-780 on exoplanets from the Ministry of Education and Science.

*Translated by E. Petrova*